\def\df{f}
\def\Re{{\rm Re}}
\def\Im{{\rm Im}}
\def\pw{{\rm pw}}

\documentstyle{laa}

\begin{document}

\thesaurus{
	11.11.1 
	11.19.1 
	11.19.6 
          }

\title{Numerical calculation of linear modes in stellar disks.}
 
\author{P.~Vauterin\thanks{Research assistant, NFWO} \and H.~Dejonghe}

\institute{Universiteit Gent, Sterrenkundig Observatorium, 
Krijgslaan 281, B--9000 Gent, Belgium}

\date{Received date; accepted date}

\maketitle 


\begin{abstract}
We present a method for solving the two-dimensional linearized
collisionless Boltzmann equation using Fourier expansion along the
orbits. It resembles very much solutions present in the literature,
but it differs by the fact that everything is performed in coordinate
space instead of using action-angle variables. We show that this
approach, though less elegant, is both feasible and
straightforward.

This approach is then incorporated in a matrix method in order to
calculate self-consistent modes, using a set of potential-density
pairs which is obtained numerically. We investigated the stability of
some unperturbed disks having an almost flat rotation curve, an
exponential disk and a non-zero velocity dispersion. The influence of
the velocity dispersion, halo mass and anisotropy on the stability is
further discussed.

\end{abstract}

The study of the perturbations in stellar disks (spirals and bars) has
advanced along essentially 2 quite different avenues since the early
60's.

Numerical N--body simulations probably offer the most flexible tools
(e.g. \cite{Hohl}; Athanassoula \& Sellwood, 1986), they are
relatively easily set up regardless the complexity of the initial
conditions and include the description of nonlinear evolution. On the
other hand, these simulations are very time-consuming and suffer from
important, but hard to quantify, numerical noise.

A linearized self-consistent mode analysis, which explains linear
instability during the onset, was the first method used, because
initially N--body simulations were still essentially infeasible.
This approach is somewhat complementary to N--body simulations. It
heavily relies on simplifying assumptions but produces high quality
results, of course only in the regions where the assumptions are
valid.  Although the literature offers general methods
(\cite{Kalnajs3}) to calculate linear instabilities, these methods
apparently are difficult to apply in practical situations. For this
reason, several researchers have adopted further simplifications in
order to handle the equations more conveniently, such as using cold
disks with softened gravity (\cite{Toomre2}) and a gaseous
approximation (\cite{Bertin_et_al}).

In this paper we develop an alternative method (section 2, 3 and 4),
which resembles very much other approaches described in the literature
(e.g. \cite{Kalnajs3}, \cite{Hunter2}), but differs on a few important
points. As in many other cases, the linearized Boltzmann equation is
solved by fourier expansion of the perturbation along the unperturbed
orbits. However, we perform all calculations in coordinate space
instead of writing everything in action-angle variables. Despite the
fact that the equations are less elegant, this strategy offers the
advantage that the perturbed mass density as well as the full
perturbed distribution is obtained in ordinary space and velocity
coordinates. This considerably simplifies the solution of the Poisson
equation, since any complete set of potential-density pairs is now
sufficient to expand the perturbing potential. This is in
contrast to the action-angle implementations, where usually
bi-orthogonal sets enter the calculations (see Kalnajs (1977) for
more details).  In addition, the equations can be cast in such a form
that all calculations are very efficient, since the
evaluation of the response mass density is essentially reduced to the
calculation of a Hilbert transform.

We applied this method to discuss the instabilities of a family of
unperturbed disk models (described in section 1). All models feature a
reasonably flat rotation curve and an exponential disk. This family
consists of almost isotropic distribution functions, with varying
velocity dispersions. Stability increases with increasing velocity
dispersions and increasing halo mass (section 5). Section 6 sums up.


\section{The models}
Before presenting the method of the mode calculation in more detail,
we first introduce the unperturbed galaxy models for which the
calculations were performed.  All our models have the same unperturbed
potential and mass density, but feature different distribution
functions, with varying velocity dispersion, streaming velocity and
anisotropy.

\subsection{The unperturbed potential}
\begin{figure}
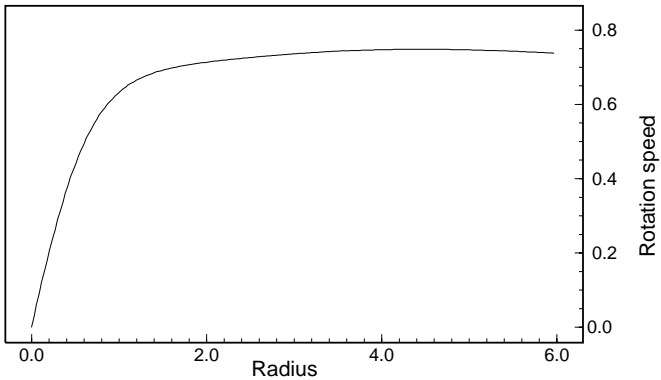

\vspace{5.5cm}
\special{ hscale=80 vscale=80 hoffset=-24 voffset= -15
hsize=480 vsize=500 angle=0 psfile=4583F1.ps } 
\caption{Rotation curve of the unperturbed \label{figrot} galaxy potential}
\end{figure}

The potential of the unperturbed galaxy was constructed as a sum of
two Kuzmin-Toomre disk potentials with different core radii. In the
plane of the disk, it is given by
\begin{equation}
V_0(r)={1 \over \sqrt{1+r^2}} + {1 \over \sqrt{1+(r/4.4)^2}}. \label{pot0}
\end{equation}
Note that potentials are defined as binding energies, with a positive
sign. This potential produces a rotation curve which is much flatter
than a single component Kuzmin-Toomre potential (see fig.
\ref{figrot}). The ratio of the flat part to the rising part is about
$6/1$.

Although the rotation curve tends to rise somewhat too slowly
near the centre, it has, at least qualitatively, a realistic
behaviour.

\subsection{The mass density profile}
It is well-known that strongly flattened galaxies usually have a
substantial amount of dark halo mass, extending much further than the
visible component. It is sufficient for our purpose to model it by a
spherical and pressure-supported galaxy component.  Therefore it is
reasonable to assume that this halo component only influences the
stability behaviour of the disk by its contribution to the global
potential. The same roughly holds for the central bulge, which is hot
and has a three-dimensional structure as well. Thus both the halo and
the bulge are ``inert'' and the corresponding potential and mass
density are taken to be spherical and denoted by $V_{0,H}(r)$ and
$\rho_{0,H}(r)$.

The disk itself is the only component which is supposed to consist of
``active mass'', sensitive to instabilities. The unperturbed disk is
supposed to be two-dimensional and axisymmetric, with a surface
density $\rho_{0,D}$ and a potential $V_{0,D}$.  We chose an
exponential mass profile with a central core:
\begin{equation}
\rho_{0,D}=\alpha e^{-1.3 \sqrt{0.2+r^2}} \label{dens0}.
\end{equation}
Although this mass density extends up to infinity, the actual models
have an outer limit at $r=6$, and the mass density reaches zero at
that point. This we achieve by fitting the distributions (\ref{dens0})
with several finite components. The relative error made at the outer
edge is only of the order of 0.1\%. The parameter $\alpha$ determines
the total disk mass.

Since the total system should be self-consistent, the total potential
in the plane of the galaxy is given by the sum of the disk and halo
component:
\begin{equation}
V_0(r)=V_{0,D}(r)+V_{0,H}(r).
\end{equation}
Since the total potential $V_0$ has a fixed form (\ref{pot0}) and the
potential of the disk is determined by the surface mass density
(\ref{dens0}), we can calculate $V_{0,H}=V_0-V_{0,D}$.  Thus follows
the halo mass density
\begin{equation}
\rho_{0,H}= -{1 \over 4 \pi G} {1 \over r^2} {d \over dr} 
  \left( r^2 {d V_{0,H} \over dr} \right),
\end{equation}
and the total halo mass within a radius $r$
\begin{equation}
M_{0,H}(r)= -{1 \over G} r^2 {d V_{0,H} \over dr}.
\end{equation}

\begin{figure}
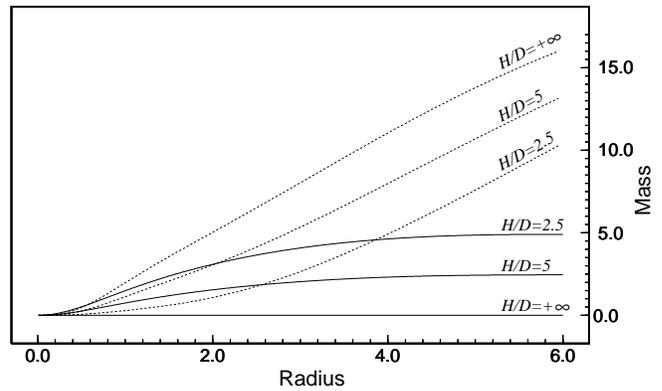

\vspace{5.5cm}
\special{ hscale=80 vscale=80 hoffset=-62 voffset=-30
hsize=480 vsize=500 angle=0 psfile=4583F2.ps }
\caption{Total mass inside $r$
of the disk (full curves) and the halo (dashed curves) for $H/D$=$2.5$,
$5$ and $+\infty$.}
\end{figure}

Of course, the halo mass density should everywhere be positive. This
puts an upper limit on the value of $\alpha$ in (\ref{dens0}). The
relative contribution of the disk and halo components are quantified
using the halo-to-disk factor, $H/D$, which gives the proportion of
the total mass inside the radius $r_{\rm max}$ (=6) for the halo and the
active disk. Self-consistency with a non-negative spherical halo puts
a lower limit on $H/D$ of about $2.5$. Note that the assumption of a
spherical halo is not a crucial one. If one would assume an oblate
halo, the only effect would be a somewhat lower minimum $H/D$.

\subsection{The distribution function}
We will examine the stability behaviour for different stellar
distributions. According to Jeans' theorem, the unperturbed part of
the distribution is a function of two integrals of motion, the binding
energy $E$ and the angular momentum $J$, defined by
\begin{equation}
E=V_0(r)-{1 \over 2} ( v_r^2+v_\theta^2)
{\rm \ \ \  and \ \ \ }
J=r v_\theta.
\end{equation}
In order to generate a variety of finite disks, based on the potential
(\ref{pot0}), the distribution function is written as a linear
combination of basic distributions:
\begin{equation}
\df_0(E,J)=\sum_{t=1}^{n_t} c_t \df_{0,t}(E,J).
\end{equation}
All components only have a (everywhere positive) contribution for
orbits lying completely inside $r_{\rm max}$, i.e. for the region
where (see also fig \ref{lind1})
\begin{equation}
E \ge V_0(r_{\rm max}) - {J^2 \over 2 r_{\rm max}^2},
\end{equation}
\begin{equation}
E \ge V_0(r_{\rm max})-{1 \over 2} v_{circ}^2(r_{\rm
max}),\label{excl2}
\end{equation}
with $r_{\rm max}$ the radius of the edge of the disk. Equation
 (\ref{excl2}) is required in order to exclude the region $r_-
\ge r_{\rm max}$. In addition, the distribution goes to
zero at this edge in a smooth way, so that the first derivative
remains finite everywhere.  The explicit form of the components is
listed in the Appendix.

The expansion coefficients $c_t$ are determined by a least square fit
of the corresponding mass density to the proposed exponential form
(\ref{dens0}) (the coefficients are forced to be positive, in order to
avoid negative distributions). By choosing an appropriate set of
components $\df_{0,t}$, we were able to create the desired orbital
densities. In all the models, the error on the fit to the mass density
never exceeds 1\% of the central value.

\begin{figure}
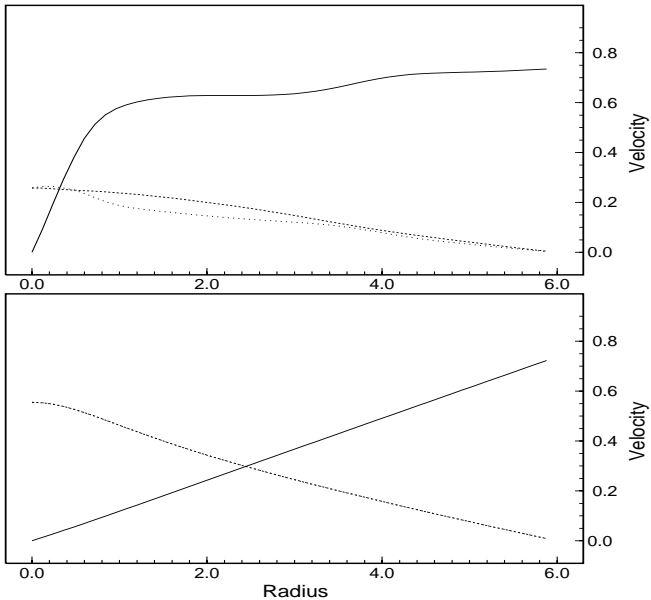

\vspace{8.0cm}
\special{ hscale=80 vscale=64 hoffset=-22 voffset= -17
hsize=480 vsize=500 angle=0 psfile=4583F3.ps }
\caption{Streaming \label{figvelocI}
velocity (full line), azimuthal velocity dispersion (long dashed line)
and radial velocity dispersion (short dashed line) for Model IV (top
panel) and Model I (bottom panel).}
\end{figure}

We constructed 4 models, labeled I to IV (the explicit form of the
distribution functions is listed in the appendix). Along the sequence,
the models become more and more rotation-supported, having an
increasing streaming velocity and decreasing temperature. Fig.
\ref{figvelocI} shows the streaming velocities and dispersions for the
coldest and hottest case. The dispersions all go to zero at the edge
of the disk. Note that model I is perfectly isotropic and has a
linearly increasing mean velocity curve. For all disks, Toomre's local
axisymmetric instability criterion (\cite{Toomre64})
\begin{equation}
Q= { \sigma_r \kappa \over 3.36 G \rho_{0,D} } \label{defQ}
\end{equation}
(with $\sigma_r$ the radial velocity dispersion and $\kappa$ the
epicyclic frequency) is everywhere higher than $1$.

\begin{figure}
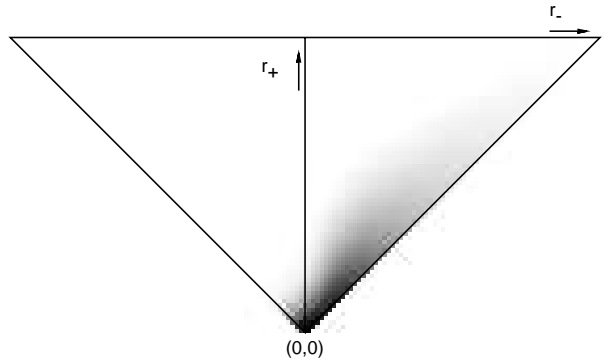

\vspace{5.0cm}
\special{ hscale=70 vscale=70 hoffset=-17 voffset= -10
hsize=480 vsize=500 angle=0 psfile=4583F4.ps }
\caption{Turning point \label{figdf0}
representation of the distribution function of model III.}
\end{figure}

In fig. \ref{figdf0}, the distribution function of model III is shown
in turning point space. The turning points of an orbit are the largest
(resp. smallest) distance from the centre that an orbit can reach, and
are called apocentre $r_+$, or pericentre $r_-$.  By convention, we
take $r_+$ always positive, while $r_-$ has the same sign as $J$.
Evidently, these quantities are integrals of the motion and $r_+ \ge |
r_- |$ (on circular orbits, $r_+=r_-$).  For a given pair $(r_+,r_-)$,
the energy and angular momentum follow immediately from
\begin{equation}
E=V_0(r_{+,-})-{1 \over 2} { J^2 \over r_{+,-}^2} \label{Eext},
\end{equation}
which leads to
\begin{equation}
E={r_+^2 V_0(r_+)    -     r_-^2 V_0(r_-) \over r_+^2 - r_-^2}
\end{equation}
and
\begin{equation}
J=\sqrt{2}r_+ r_- \sqrt{V_0(r_+)-V_0(r_-) \over r_-^2 - r_+^2}.
\end{equation}
Inversely, $r_+$ and $r_-$ are found as the roots for $r$ of
(\ref{Eext}) for a particular $E$ and $J$.  We preferred these
variables over the normal $(E,J)$ space, not only because they have an
easy physical interpretation, but also because this representation is
more related to our method for solving the linearized Boltzmann
equation, which employs a grid interpolation in the
turning point space (see following section).

We have chosen finite disks in order to compactify phase space.
However, as can be seen from the distribution function (fig.
\ref{figdf0}), the disk reaches this limit at $r_{\rm max}$ in a very
smooth way. This is important since it has been proven that a sharp
edge or, more generally, a sharp feature in $(E,J)$ space, can introduce
additional instabilities which might not always be physical
(\cite{Toomre64}, \cite{Sellwood_Kahn}).

\section{The Poisson equation}

\begin{figure*}
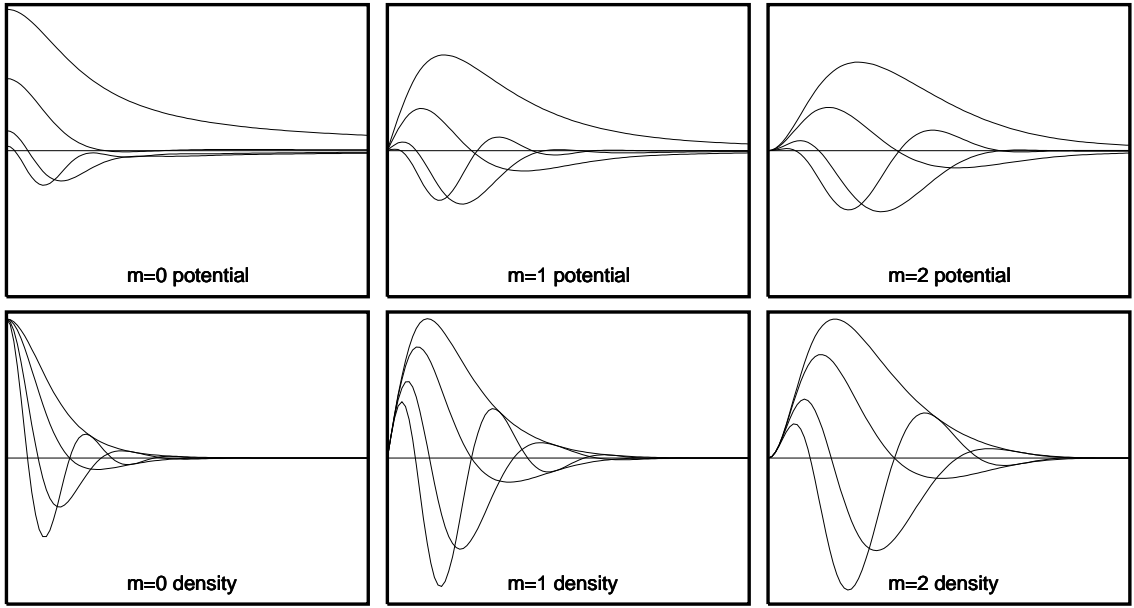

\vspace{8.0cm}
\special{ hscale=90 vscale=90 hoffset=30 voffset=-10
hsize=480 vsize=500 angle=0 psfile=4583F5.ps }
\caption{Some of the lowest order potential-density pairs used to
expand a general perturbation.}
\end{figure*}

Since we want to study uniformly rotating perturbation modes, we
suppose the following form for the perturbing potential:
\begin{equation}
V'(r,\theta;t)=V'(r) e^{i(m \theta - \omega t)}.
\end{equation}
In this expression (and in all later similar expressions), only the
real part corresponds to the physical quantity. There is no problem
though to calculate with these complex expressions, since all
equations are linear.  This perturbation is rotating with a pattern
speed of $\Re(\omega)/m$ and is exponentially growing with a growth
factor $\Im(\omega)$. A more general pattern can be obtained as a
superposition of such components with different $m$.

We will search for the linear modes using a matrix method, which
requires a set of potential-density pairs suitable for the expansion
of the perturbation. The literature offers a variety of analytical
sets for infinite disks (eg. \cite{Clutton_Brock},
\cite{Qian}) as well as for finite ones (eg. \cite{Hunter}).
The choice of a particular set largely depends on the structure of the
unperturbed potential $V_0$ and density $\rho_0$. The importance of a
suitable set of potential-densities is clearly illustrated by the
displacement mode of a self-consistent disk. If the model has no inert
halo component, a simple translation of the system is a valid
``perturbation''. In linear theory, this corresponds to an $m=1$ mode
of the form (see also section \ref{verplaatstest})
\begin{equation}
{d \rho_0 \over d r} (r) e^{i \theta}.
\end{equation}
The potential-density pairs should of course be able to fit this
behaviour well (\cite{Weinberg}).  Moreover, it is very important to
choose the set as efficiently as possible, so that not too 
many expansion terms need to be calculated.  For these reasons, we
decided not to rely on existing analytical sets, but to create
numerical sets which are tailor-made for the present
unperturbed disk.  If the maximum radius of the finite disk is denoted
by $r_{\rm max}$, the set of densities for $m=0$ is given by
\begin{equation}
\rho^0_{S,n}(r)=\rho_{0,D}(r) *
	      \cos(\pi n {r \over r_{\rm max}}),
\end{equation}
and for $m>0$ by
\begin{equation}
\rho^m_{S,n}(r)=r^{m-1} {d \rho_{0,D} \over d r} (r) * 
	      \cos(\pi n {r \over r_{\rm max}}).
\end{equation}
These densities have the right behaviour in the centre and at the edge
of the galaxy. In addition, their number of radial nodes is equal to
$n$. The corresponding potentials $V^m_{S,n}$ are easily calculated
numerically using the integral form of the Poisson equation. These
potential-density pairs prove to be very efficient for the expansion
of the normal modes. Of course, these modes in principle could have
been expanded using analytical complete sets described in the
literature (e.g. \cite{Hunter}), but this would require a long
summation up to very high order, particularly because the density
falls to zero very smoothly at the edge.

\section{The linearized Boltzmann equation}
\subsection{Integral form}
The total disk is modeled using two parts, a time-independent
axisymmetric unperturbed part, and a small perturbation. Hence, the
distribution function is written as
\begin{equation}
\df(r,\theta,v_r,v_\theta;t)=\df_0(E,J)
	  +\df'(r,\theta,v_r,v_\theta;t),
\end{equation}
while the total potential, which is defined as a binding energy (being
positive everywhere), reads
\begin{equation}
V(r,\theta;t)=V_0(r)+V'(r,\theta;t).
\end{equation}
If the system is exposed to the perturbing potential $V'$, the
corresponding linearized response distribution is given by
(\cite{Vau_Dej})
\begin{equation}
\df'={\partial \df_0 \over \partial E} \df'_E +
     {\partial \df_0 \over \partial J} \df'_J,\label{simpsol}
\end{equation}
where $\df'_E$ and $\df'_J$ satisfy (using Poisson brackets)
\begin{equation}
{\partial \df'_E \over \partial t}-
[\df'_E, E]= [E,V']
\end{equation}
and
\begin{equation}
{\partial \df'_J \over \partial t}-
[\df'_J, E]= [J,V'].
\end{equation}
Integration with respect to the time along unperturbed orbits of both
equations and expansion of the Poisson brackets immediately yields
(since the left hand side is the total time derivative along an
unperturbed orbit)
\begin{eqnarray}
\df'_E(r^0,\theta^0,v_r^0,v_\theta^0;0) &=& \int_{-\infty}^0
  (v_r {\partial V' \over \partial r} 
 + {v_\theta \over r} {\partial V' \over \partial \theta}
  ) dt \\
\df'_J(r^0,\theta^0,v_r^0,v_\theta^0;0) &=& - \int_{-\infty}^0
  {\partial V' \over \partial \theta} dt.
\end{eqnarray}
The integrand is to be evaluated along the unperturbed orbit which
passes at $t=0$ through the point
$(r^0,\theta^0,v_r^0,v_\theta^0)$. One should note that these
expressions can only hold if the perturbation was zero at $t=-\infty$
and is, at least infinitesimally, growing.

With a potential of the form
\begin{equation}
V'(r,\theta,t)=V'(r) e^{i(m \theta - \omega t)},
\end{equation}
we have
\begin{eqnarray}
\df'_E&=& \int_{-\infty}^0 (v_r {d V' \over dr} + i m v_\theta {V' \over r})
	       e^{i(m \theta-\omega t)} 
         dt \label{dfE1} \\
\df'_J&=& \int_{-\infty}^0 i m V' e^{i(m \theta - \omega t)}
         dt \label{dfJ1} .
\end{eqnarray}
These integrals are to be evaluated along unperturbed orbits, so
one needs a way to handle such orbits.

\subsection{Fourier expansion along unperturbed orbits}
Since the unperturbed potential $V_0$ is axisymmetric, the
corresponding orbital structure is particularly simple. Taking
advantage of the conservation of angular momentum $J$, one can
conclude that the radial coordinate behaves like the coordinate of a
particle in a one-dimensional (so-called effective) potential
\begin{equation}
V_{\rm eff}(r)=V_0(r)-{J^2 \over 2 r^2}.
\end{equation}
Supposed that the orbit is bound (which is the only case of interest
for galaxies), the radial coordinate along an unperturbed orbit should
be a periodic function of time, with angular frequency $\omega_r$.
From this it follows immediately that $v_r$ and $v_\theta$ are periodic
with frequency $\omega_r$ as well. As the mean value of $v_\theta$
should not necessarily be zero, the angular coordinate $\theta$ will
be a superposition of a periodic function and a uniform ``drift''
velocity:
\begin{equation}
\theta=\omega_\theta t + \theta_p(t)
\end{equation}
(with $\theta_p$ a periodic function with period $\omega_r$).

We will now rewrite the integrands of (\ref{dfE1}) and (\ref{dfJ1}) by
factorizing the part which is periodical with angular frequency
$\omega_r$:
\begin{eqnarray}
\df'_E&=& \int_{-\infty}^0 I_E(t) e^{i(m \omega_\theta-\omega )t} 
   dt \label{dfE2} \\
\df'_J&=& \int_{-\infty}^0 I_J(t) e^{i(m \omega_\theta-\omega )t} 
   dt \label{dfJ2} ,
\end{eqnarray}
with
\begin{equation}
I_E(t)=\left( v_r(t) {\partial V' \over \partial r}(t) 
	+ i m v_\theta(t) {V'(t) \over r(t)} \right) e^{i m \theta_p(t)}
\end{equation}
and
\begin{equation}
I_J(t)= i m  V'(t) e^{i m \theta_p(t)}.
\end{equation}
Since $I_E(t)$ and $I_J(t)$ are periodic, they can be expanded in
Fourier series:
\begin{eqnarray}
I_E(t)= \sum_{l=-l_{\rm max}}^{l_{\rm max}} I_{E,l} e^{i l \omega_r t} \\
I_J(t)= \sum_{l=-l_{\rm max}}^{l_{\rm max}} I_{J,l} e^{i l \omega_r t}
\end{eqnarray}
When these forms are substituted in (\ref{dfE2}) and (\ref{dfJ2}), the
integrations can be carried out analytically, at least if the
perturbation is growing, $\Im(\omega)>0$. Note that, when the
perturbation is decaying in time ($\Im(\omega)<0$), one can still
obtain a solution by performing the integrations (\ref{dfE2}) and
(\ref{dfJ2}) from $t=0$ to $t=+\infty$.

\subsection{Grid interpolation}
The Fourier expansion strategy yields the response distribution for
any given perturbing potential, but a number of steps are involved
even for calculating one single point of the distribution: the orbit
should be integrated for at least one half of the radial period, and
the appropriate functions are to be expanded in Fourier series. Since
the perturbed distribution will be evaluated many times (e.g. for the
calculation of the perturbed mass density one has to integrate over
the velocities), it is absolutely necessary to be able to evaluate the
response as fast as possible. Our solution is to calculate all
appropriate parameters (the frequencies $\omega_r$ and $\omega_\theta$
and the Fourier expansions $I_{E,l}$ and $I_{J,l}$) on a
(two-dimensional) grid in integral space and to store the result on
disk for future use. We found that is was not so convenient to use
$E,J$ as grid coordinates, since this grid is very inhomogeneous
and has a relatively low density near the circular orbits, where
disk galaxies have a dense population. Moreover, the circular orbit
limit on this grid is usually not given in an analytic form. For these
reasons, we used $r_+$ and $x=r_-/r_+$ as grid coordinates. In this
representation, we can apply a simple rectangular grid which is very
dense close to circular orbits and in the neighbourhood of the centre
of the galaxy.

The $(r_+,x)$ space is, up to the maximum radius $r_+=r_{\rm max}$,
discretized using a rectangular grid. On each point of this grid, the
Fourier expansion for the orbit starting at $t=0$ in its apocentre (by
choice) is calculated and stored in a library, together with a
tabulation of $t[r]$ and $\theta_p[r]$. In the following
paragraphs, we will frequently use both the time and $r$--dependence
of the same variable. In order to avoid confusing notations,
$r$--dependences will be written using square brackets.
The time and coordinate system is chosen in such a way that $\theta$
and $t$ are zero at the apocentre.

When the response distribution is later required in a point $p^0=(r^0,
\theta^0=0, v_r^0, v_\theta^0; t^0=0)$ in phase space, the
corresponding turning points $r_+^0$ and $r_-^0$ are calculated and
all parameters are interpolated using the four closest points on the
grid (note that, since the response is a known periodic
function of $\theta$, it is sufficient to evaluate it at $\theta=0$).
In this way, we get an orbit out of the library with the correct
integrals of motion, but passing through its apocentre at $t=0$. We
will denote this library orbit and all corresponding quantities with a
superindex $L$.

\begin{figure}
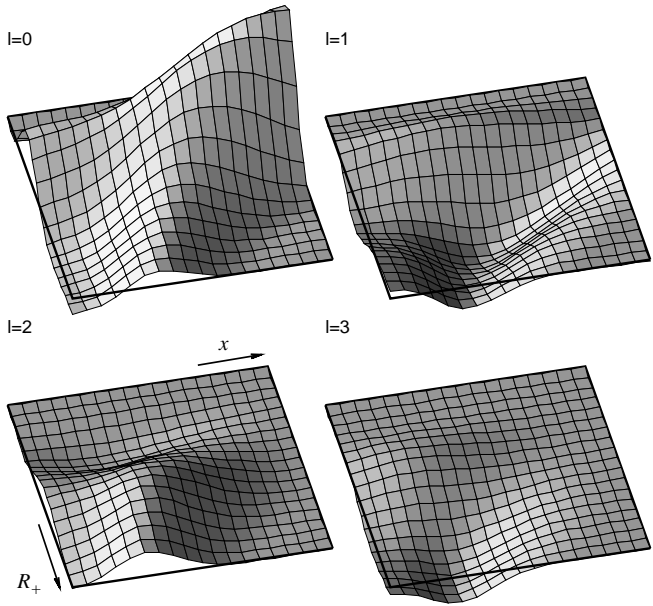

\vspace{9.0cm}
\special{ hscale=80 vscale=80 hoffset=-10 voffset=-10
hsize=480 vsize=500 angle=0 psfile=4583F6.ps }
\caption{Interpolation grids of the lowest  \label{figfour} 
order Fourier coefficients for a $V_{S,1}^2$ perturbation (only the
$x>0$ part is shown)}
\end{figure}

As the point $p^0$ is in general not the apocentre of the orbit, one
should take in account an offset in time for the actual orbit passing
through $p^0$:
\begin{equation}
I_E(t)=I_E^L ( t+t^L[r^0] ),
\end{equation}
Note that $t^L[r^0]$ is actually a double--valued relation.  However,
the sign of $v_r^0$ can be used to determine which branch of the
relation should be used. In general, there is also an offset in
azimuth:
\begin{equation}
\theta(t)= \omega_\theta^L t + 
      \theta_p^L ( t+t^L[r^0] )
     -\theta_p^L[r^0],
\end{equation}
which is of course the reason why the functions $t[r]$ and
$\theta_p[r]$ are stored. Note that the presence of these functions
does not reduce the speed of the calculations, but implies only a
slight increase in used disk space. The resulting perturbed
distribution in that point follows then immediately:
\begin{equation}
\df'_E= \sum_{l=-l_{\rm max}}^{l_{\rm max}} { I_{E,l}^L 
             e^{i(l \omega_r^L t^L[r^0] - m \theta_p^L[r^0] )}
                    \over
               i ( l \omega_r^L + m \omega_\theta^L - \omega ) }
\end{equation}
and
\begin{equation}
\df'_J= \sum_{l=-l_{\rm max}}^{l_{\rm max}} { I_{J,l}^L
             e^{i(l \omega_r^L t^L[r^0] - m \theta_p^L[r^0] )}
                    \over
               i ( l \omega_r^L + m \omega_\theta^L - \omega ) }.
\end{equation}

The Fourier coefficients which are stored on the grid only depend on
the unperturbed and the perturbing potential. Furthermore, since the
latter will be expanded in a basis, this has to be done only for the
discrete set $V_{S,n}^m$. The orbits are obtained by integrating the
system over half a period using a fourth order Runge-Kutta. Special
attention should be paid to $r_-=0$ (or $J=0$) orbits, which have to
be integrated in rectangular coordinates. In addition, since the
Fourier coefficients are discontinuous at $r_-=0$ (see also
\cite{Kalnajs3}), this should be performed twice, for
``infinitesimally small'' opposite values of $J$ in order to get the
correct left and right limit.

The Fourier expansion is performed up to the order $l_{\rm max}=14$.
This number should be relatively large in order to provide an
accurate fit to the high excentricity orbits, which show fast
variations close to the centre. A grid resolution of $60
\times 60$ proved to provide more than enough accuracy,
particularly because the functions do not show very steep
gradients in the turning point space. In these circumstances and
for a set of 8 perturbing potentials, the calculations take less than
4 hours on a 100 Specfp92 workstation.  This work only has to be
redone when one changes the unperturbed potential.

Fig. \ref{figfour} shows some of the Fourier coefficient grids for a
$m=2$ perturbing potential. It clearly illustrates that the higher
order Fourier terms are only important for small $x=r_-/r_+$ and for
large $r_+$, which means for very eccentric orbits. This indicates
that, for relatively cool galaxies, a small number of Fourier terms
can already yield reliable results.

\begin{figure*}
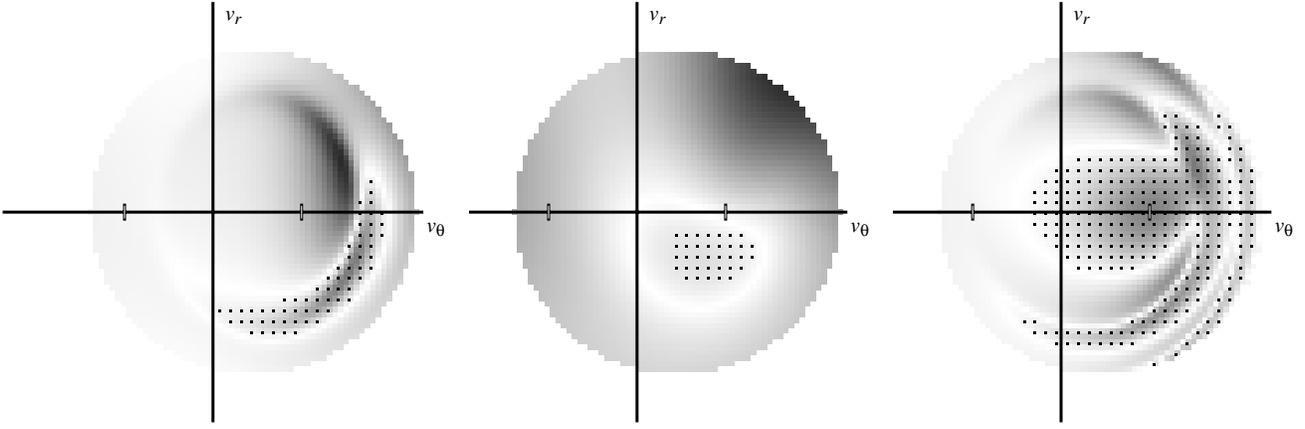

\vspace{6.0cm}
\special{ hscale=94 vscale=94 hoffset=-15 voffset= -23
hsize=530 vsize=500 angle=0 psfile=4583F7.ps }
\caption{The real part of the perturbed distribution 
$\Re(\df'_E)$ at $r^0=1.8$ for a typical
\label{figpdf} $m=2$ perturbation. From left to right:
at $\omega=(0.7,0.07)$, $\omega=(0.7,0.7)$ and $\omega=(1.5,0.1)$
(dotted regions are negative responses; the small ticks on the
$v_\theta$ axis correspond to circular velocities).}
\end{figure*}

Fig. \ref{figpdf} displays some response distribution functions for
different $\omega$. The left and middle panels have the same pattern
speed $\Re(\omega)/m$ and are taken somewhat inside the corotation
radius.  The left panel shows a response with a much smaller growth
rate and therefore has a more prominent corotation resonance,
appearing as a ring-shaped feature. This picture still has to be
multiplied with $\partial \df_0 / \partial E$ which, for relatively
cool disks, will emphasize the stars in the neighbourhood of circular
orbits. One of the most crucial effects of non-zero velocity
dispersion is that the distribution function populates stars on all
possible orbits that are resonant with the perturbations, not only the
resonances on circular orbits. These orbits fill a complicated region
in phase space, cuts of which can be seen in velocity space in fig.
\ref{figpdf}.  Due to the differential rotation, a substantial
number of orbits at this radius are in corotation resonance,
although this is not the corotation radius.

\begin{figure}
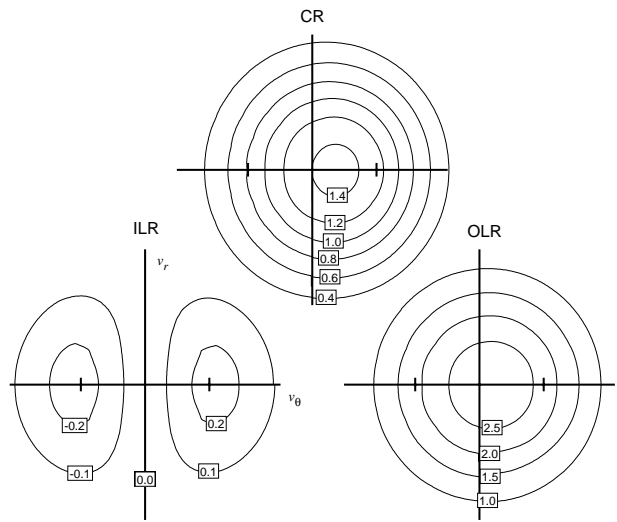

\vspace{7.0cm}
\special{ hscale=90 vscale=90 hoffset=-10 voffset= -20
hsize=480 vsize=500 angle=0 psfile=4583F8.ps }
\caption{The resonant values of  \label{figreso} 
$\omega$ at $r=1$ in the $(v_\theta,v_r)$ plane for the three most
important $m=2$ resonances. Circular velocities are indicated.}
\end{figure}

Fig. \ref{figreso} shows, for a fixed radius, the velocity dependence
of the resonant $\omega$ value for the dominant $m=2$ resonances. This
value is given by
\begin{equation}
\omega= l \omega_r+m \omega_\theta. \label{defreso}
\end{equation}
The lowest order resonances for stars rotating in the direct sense
($\omega_\theta>0$) are $l$= $-1$ (ILR), $0$ (CR) and $1$ (OLR).
However, there is a problem in this definition for counterrotating
stars, since $\omega_\theta$ discontinuously changes its sign at
$v_\theta=0$. This can be resolved by noticing that if, for
counterrotating stars, $l$ is replaced by $l+m$ in (\ref{defreso}),
the resulting $\omega$ has an overall smooth behaviour. This is a
consequence of the fact that for zero momentum stars $\omega_r=2
\omega_\theta$. The ring-shaped CR position, as shown in fig. \ref{figreso}
(for varying $\omega$), can easily be found back in the left panel of
fig. \ref{figpdf} (for a specific value of $\omega$).

Since $\omega_r$ and $\omega_\theta$ depend mostly on the energy $E$,
the CR, OLR and higher resonances are roughly concentric circles
around the origin. At the ILR though, the sum largely cancels out and
result is mostly dominated by the angular momentum $J$.

\subsection{Integration over the velocities}
The method of the previous section can be used to calculate the
perturbed distribution in a particular point of phase space
$(r,\theta=0,v_r,v_\theta)$. Writing the velocities $(v_r,v_\theta)$
in polar coordinates $(v,\alpha)$, this results in a response
\begin{equation}
\df'(r,\theta=0,v,\alpha,t=0)=
\sum_{l=-{l_{\rm max}}}^{l_{\rm max}} {A_l(r,v,\alpha) 
   \over p_l(r,v,\alpha) - \omega},
\end{equation}
with
\begin{equation}
p_l(r,v,\alpha)=l \omega_r + m \omega_\theta.
\end{equation}
In order to obtain the perturbed mass density at a radius $r$ and for
$\theta=0$, one has to integrate this expression over the velocities
up to the escape velocity. Since the position of the pole $p_l$ in
general depends also on the velocities, the denominator should be kept
inside this integration:
\begin{equation}
\rho'(r)= \sum_{l=-{l_{\rm max}}}^{l_{\rm max}} \int_0^{v_{esc}(r)} \int_0^{2\pi}
{A_l(r,v,\alpha) \over p_l(r,v,\alpha) - \omega} v dv d\alpha. \label{rhoaa}
\end{equation}

In each term of the sum, we will switch the integration variables from
$(v,\alpha)$ to $(p_l(r,v,\alpha),\alpha)$. The velocity then becomes
a function $v=V_l(r,p_l,\alpha)$ and the integral is written as
\begin{eqnarray}
\lefteqn{
\rho'(r)=\sum_{l=-{l_{\rm max}}}^{l_{\rm max}} \int_{p_{l,min}}^{p_{l,max}}
	 {1 \over p_l-\omega} d p_l
\nonumber \times 
} \\
\lefteqn{
  \int_0^{2\pi} A_l(r,V_l(r,p_l,\alpha),\alpha) V_l(r,p_l,\alpha)
  {\partial v \over \partial p_l} (r,p_l,\alpha) d \alpha. \label{vsi}
}
\end{eqnarray}
The limits of the first integral can be chosen freely, as long as the
interval is large enough to cover all poles.  If $p_l(r,v,\alpha)$ is
not a monotone function of $v$ over the entire region of interest,
this transformation can still be performed by splitting up the
integral in parts. There is a simple reason to write the expression in
this way: for a constant $r$, the inner integrals are functions which
only depend on $p_l$.  If we sum all these functions over $l$ and
denote this summation with $W(r,p)$, the perturbed mass density
reduces to
\begin{equation}
\rho'(r)=\int_{p_{\rm min}}^{p_{\rm max}} {W(r,p) \over p-\omega}
   dp, \label{intrho}
\end{equation}
which is a simple weighted integral over all real pole positions.
This integral represents the Hilbert transform of $W$ with respect to
$p$, assuming that $\rho$ is also a function of $\omega$. For each
value of $r$, the function $W(r,p)$ can be tabulated and stored for
later use. The evaluation of the perturbed density for any value of
$\omega$ is then reduced to the numerical evaluation of a
one-dimensional integral, which can be performed very fast.

\begin{figure}
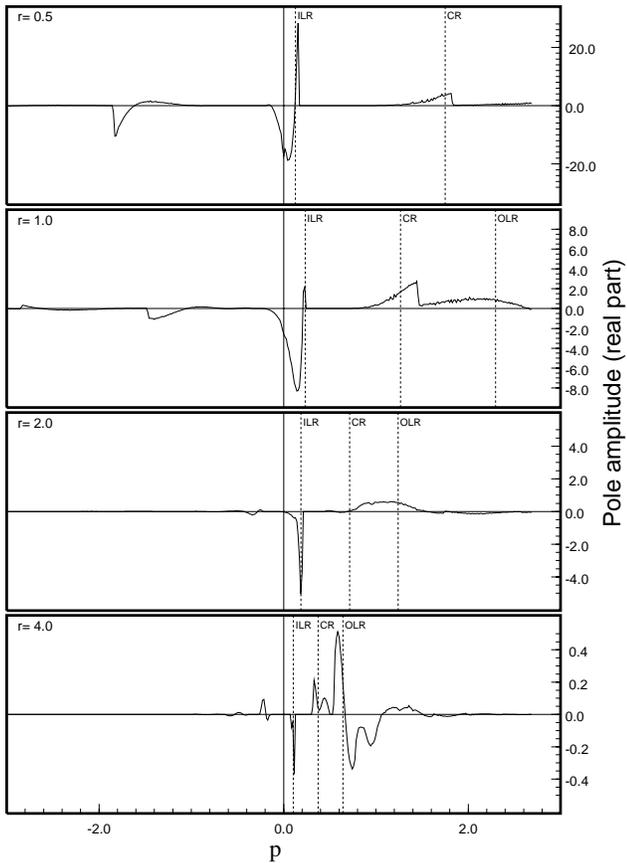

\vspace{12.0cm}
\special{ hscale=90 vscale=90 hoffset= -15 voffset= -20
hsize=480 vsize=500 angle=0 psfile=4583F9.ps }
\caption{The pole distribution $W(\omega)$ for the
perturbed mass \label{figpoles} density using model III and a
$V_{s,0}^2$ perturbation. The bumps at negative $p$ correspond to a
counterrotating resonance.}
\end{figure}

In practice, $W(r,p)$ is not calculated using the explicit
integral (\ref{vsi}), because this would require the (piecewise)
inversion of $p_l(v)$, which can be awkward. Instead, an alternative
method is used, based on the two-dimensional integral (\ref{rhoaa}).
For any value of $r$, $W(p)$ is maintained in a tabular form,
containing function values at fixed points with interleave $\Delta p$.

A numerical integration of the two-dimensional integral (\ref{rhoaa})
is performed, using a rectangular grid over $[0,v_{\rm esc}] \times
[0,2 \pi]$, with interleaves $\Delta v$ and $\Delta
\alpha$. In each cell of this grid, the integrand of (\ref{rhoaa}) is
approximately constant with respect to $v$ and $\alpha$. Therefore,
the contribution of this cell to the integral is a pole function with
respect to $\omega$, having a pole position $P=p_l(r,v,\alpha)$ and an
amplitude $I=A_l(r,v,\alpha) v \Delta v \Delta
\alpha$. This amplitude is now added to the table value of $W(p)$
that has a $p$ value lying closest to $P$. When all cells are summed,
dividing all values in the $W(p)$ table by $\Delta p$ yields a
numerical tabulation of $W(p)$.  In practice, the summation over the
cells is refined using a trapezoidal rule.

The values of the function $W(r,p)$ are stored on a grid containing 40
points in the $r$ dimension and $10^4$ points in the $p$ dimension,
and the integral (\ref{intrho}) is calculated by linear interpolation
between the points in the $p$ dimension. However, one should be
careful with values of $\omega$ lying in the neighbourhood of the
real axis. If $\Im(\omega)$ is zero, the result is meaningless when
$W(r,p)\ne0$ in a region $[\Re(p)-\epsilon,\Re(p)+\epsilon]$ (in this
case, stars are in resonance with a steady perturbation and the linear
approximation breaks down anyhow). In addition, for such a region of
non-zero weight for poles, $|\Im(\omega)|$ should be large in
comparison with the grid distance of the $p$ grid in order to have a
reliable interpolation. This is the reason why we take such a
dense grid for $p$.

In fig. \ref{figpoles}, the pole distribution $W(r,p)$ at various
radii for a typical configuration is shown. For small $r$, the
contributions of each resonance are easily distinguished, while for
large $r$ the behaviour becomes more complex since the stars move
slowly and many higher order resonances occur. This picture also
clearly illustrates the non-localized character of the resonances,
particularly for the corotation (CR) and outer Lindblad resonance
(OLR).  The inner Lindblad resonance (ILR) is, even for hot disks,
much sharper, and remains roughly at the same value for $\omega$
throughout the galaxy.  As noticed earlier (\cite{Lin}), this
interesting behaviour follows from the structure of the unperturbed
potential. Due to this behaviour, the ILR plays a crucial role in the
stability behaviour, even for high velocity dispersions.

Note that, for the panel corresponding with $r=1$, one can easily
correlate the regions in $p$ spanned by the contributions of the
individual resonances with the variation in $\omega$ shown by the
corresponding contour plot in fig. \ref{figreso}.

\section{Construction of normal modes}
A matrix method is applied for searching the normal modes
(\cite{Kalnajs3}).  A general perturbing potential is written as
\begin{equation}
V'=\sum_{i=0}^s a_i V^m_{S,i} \label{pertpot} .
\end{equation}
Of course $s$ should be large enough so that all desired modes can be
expanded accurately (the limitation on $s$ puts a limitation on
the oscillatory behaviour of the modes).

Using the methods described in the previous paragraphs, the pole
density functions $W_{S,i}^m(r,p)$ associated with the perturbing
potentials $V^m_{S,i}$ are calculated and expanded as
\begin{equation}
W_{S,i}^m(r,p)= \sum_{j=0}^s \bar c_{j,i}^m(p) \rho_{S,j}^m.
\end{equation}
This expansion is obtained using a least square fit. One can easily
obtain $\bar c_{j,i}^m(p)$ in tabular form by performing the fit on
each of the rows in the $r$ direction of the $W(r,p)$ grid.  We can
now write the response density ${\rho'}^m_{S,i}$ for each of the
potentials $V^m_{S,i}$ as
\begin{equation}
{\rho'}^m_{S,i}=\sum_{j=0}^s c_{j,i}^m(\omega) \rho^m_{S,j},
\end{equation}
if we define
\begin{equation}
c_{j,i}^m(\omega)= \int_{p_{min}}^{p_{\rm max}}
  { \bar c_{j,i}^m(p) \over \omega - p } dp.
\end{equation}

The response potential of (\ref{pertpot}) now follows immediately:
\begin{equation}
\sum_{j=0}^s \left( \sum_{i=0}^s c_{j,i}(\omega) a_i \right) V^m_{S,j}.
\end{equation}
For a self-consistent mode, this response should be equal to the
original perturbing potential (\ref{pertpot}). So the search for
normal modes is reduced to the search for those values of $\omega$ for
which the matrix $C(\omega)$ (with elements $c_{q,p}$) has a unity
eigenvalue $\lambda$.  The corresponding eigenvector gives the
expansion of the normal mode.

\begin{figure}
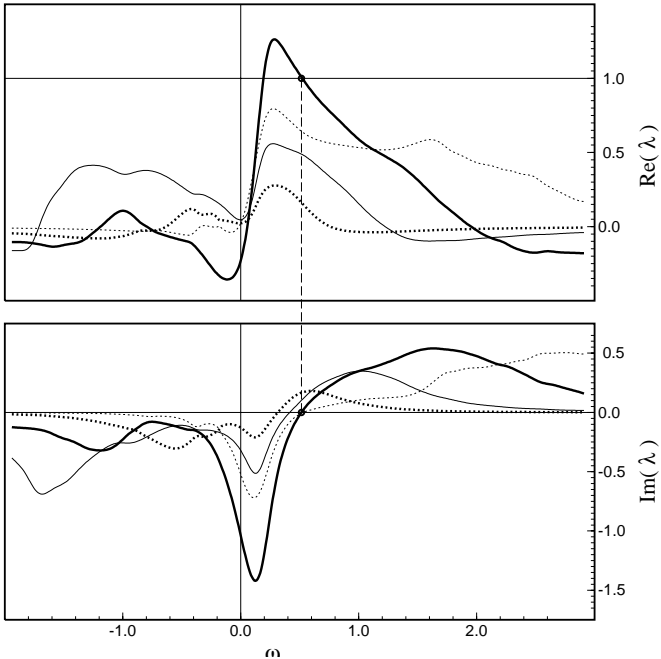

\vspace{9.0cm}
\special{ hscale=70 vscale=70 hoffset= -18 voffset= -12
hsize=480 vsize=500 angle=0 psfile=4583F10.ps }
\caption{Real and imaginary part of the \label{figeigval} multi-valued
 eigenvalue relation $\lambda(\omega)$ at $\Im(\omega)=0.11$ for the
model III. The different kinds of lines correspond to different
branches of the relation. The unique eigenvalue with $\lambda=1$ is
indicated.}
\end{figure}

A typical situation for the multi-valued eigenvalue relation
$\lambda(\omega)$ is shown in fig. \ref{figeigval}. The steep edge in
$\Re(\lambda)$ and the sharp peak in $\Im(\lambda)$ at small positive
pattern speeds are the dominating features and are caused by the ILR
because the reaction of the disk is strongest at ILR (see also fig.
\ref{figpoles}).  For each value of $\Im(\omega)$ and for each branch,
there appears to be only one (small and positive) value of
$\Re(\omega)$ where $\Im(\lambda)$ is zero. The value of
$\Re(\lambda)$, which decreases for increasing $\Im(\omega)$, further
determines the mode location.

Obviously this search for unity eigenvalues in the complex plane has
to be performed numerically. This is the main reason why we put so
much emphasis on the development of a method to evaluate $C(\omega)$
as fast as possible. It takes less than a second to calculate the
eigenvalues for a particular $\omega$, using an expansion up to an
order $s=8$. Due to the efficient structure of the
potential--density pairs, this limit is certainly sufficient for an
accurate expansion of the unstable modes, since they prove to be of a
relatively low order. The program calculates a detailed
numerical tabulation of the complete dispersion relation in a
sufficiently large region of the complex plane. This tabulation is
further used to determine all modes using a binary intersection
algorithm. This whole procedure takes less than one hour.

\subsection{Checkpoints}
Of course, there is a strong need for good checkpoints before starting
to interpret the results coming out of this method. We checked the
system in three ways, together covering more or less the whole chain
of operations. The positive results of these checks are not only a
good indication that the calculations are not obviously wrong, but
they provide a good tool to determine the values of the various
parameters (grid resolutions, expansion limits, ...) in order to get
results with sufficient accuracy. Most of these parameters
were actually determined experimentally, using these test cases.

\subsubsection{Comparison with direct integration}
For a given perturbing potential, the linearized Boltzmann equation
can be integrated numerically, at least for sufficiently fast growing
perturbations.  This yields values for the perturbed distribution
function which should be the same as those coming from the Fourier
expansion along the orbits.

\subsubsection{Uniformly rotating disks}
In a previous paper (\cite{Vau_Dej}), the mode analysis for uniformly
rotating disks has already been performed in an independent way. If
the same models are treated using the present approach, the same
results should come out.

\subsubsection{The displacement mode \label{verplaatstest} }
As mentioned already, for models without passive component, a simple
displacement of the disk is a valid ``perturbation''
(\cite{Weinberg}). In a linear approach, the responses corresponding
to this mode are obtained by derivation of the unperturbed values with
respect to a rectangular coordinate (e.g. $x$). For the unperturbed
mass density, this results in (using the fact that $x=re^{i \theta}$)
\begin{equation}
{\partial \rho_{0,D} \over \partial x}(r)= 
{\partial \rho_{0,D} \over \partial r}(r) e^{i \theta}.
\end{equation}
The distribution function gives rise to the following perturbation:
\begin{equation}
\df'={\partial \df_0 \over \partial E} {\partial E \over \partial x}
+
{\partial \df_0 \over \partial J} {\partial J \over \partial x},
\end{equation}
so that we immediately have that
\begin{equation}
\df_E'={\partial E \over \partial x}={d V_0 \over d r} e^{i \theta},
\end{equation}
and
\begin{equation}
\df_J'={\partial J \over \partial x}=v_y=(v_\theta-i v_r) e^{i \theta}.
\end{equation}
We used a self-consistent model based on a single Kuzmin-Toomre
potential to prove that the method is able to find such modes. Note
that, although this displacement mode occurs at $\omega=0$, there is
no problem with the resonances since the pole distribution $W(r,0)=0$
for $m=1$ perturbations, which we verified numerically.


\section{Results and discussion.}

We performed the linear mode analysis for the models I to IV.  These
models have a roughly isotropic distribution function, with a velocity
dispersion which is decreasing from model I to model IV.  The aim of
this set of models is to address the dependence of the stability on
two important parameters, the velocity dispersion and the halo to disk
proportion, $H/D$. Both parameters are now widely believed to have a
crucial impact on the stability. The velocity dispersion of a
particular model will be represented by its central value,
$\sigma_{\rm cent}$, which is a good estimator of the relative overall
behaviour (see fig. \ref{figvelocI}).  Unstable modes with $m$ varying
from $0$ to $4$ were  sought in all models and for varying
$H/D$. One can easily obtain the results for different $H/D$ simply by
varying the total magnitude of the unperturbed distribution function,
represented by $\alpha$ in (\ref{dens0}).

\begin{figure}
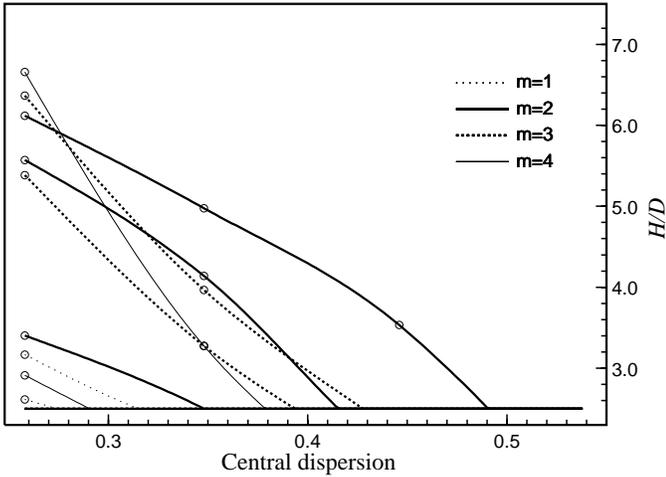

\vspace{7.0cm}
\special{ hscale=100 vscale=100 hoffset=-28 voffset= -30
hsize=480 vsize=500 angle=0 psfile=4583F11.ps }
\caption{Instability limits in the $\sigma_{\rm cent}$ vs. $H/D$ plane 
\label{figunstab} (taken as $\Im(\omega)=0.005$) for the lowest
order modes in models I to IV.  The lower left corner corresponds to
unstable regions.  The models are represented by small circles, and
the model number decreases from the left to the right (the lines
between the models are cubic spline interpolations).}
\end{figure}

\begin{figure}
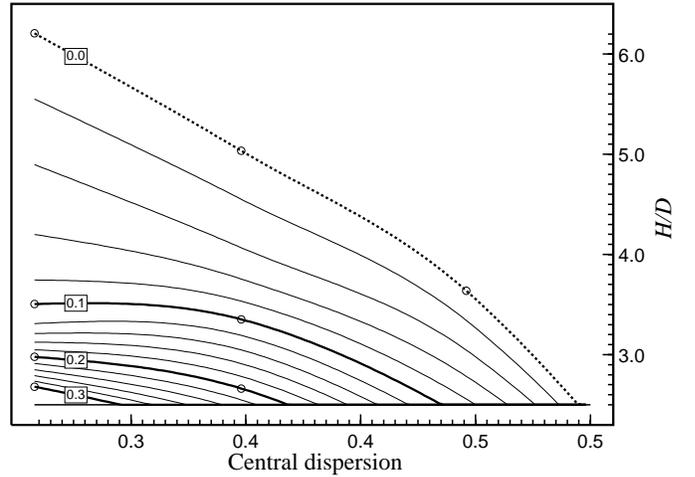

\vspace{7.0cm}
\special{ hscale=100 vscale=100 hoffset=-28 voffset= -30
hsize=480 vsize=500 angle=0 psfile=4583F12.ps }
\caption{Contour lines of the growth rate of the \label{figgrowth}
 dominating $m=2$ mode as a function of $\sigma_{\rm cent}$. The small
circles represent the model points (the lines connecting the models
are interpolations). The dotted line is an extrapolation at
$\Im(\omega)=0$, and is the same as the height $m=2$ limit in
fig. 13.}
\end{figure}

In fig. \ref{figunstab}, the stability limits for these modes are
plotted in the $\sigma_{\rm cent}$ vs.  $H/D$ plane (we chose
$\Im(\omega)=0.005$ as a stability limit, since $\Im(\omega)=0$ is
unreachable with our method).  
These curves were obtained by a cubic spline interpolation
between the four models, which are represented by small circles.  For
most of the modes, only a small fraction of the models can actually
reach the unstable regime. The other models only become unstable if
they are combined with a spherical halo which is not everywhere
positive (and hence $H/D<2.5$). The curves in figure
\ref{figunstab} and \ref{figgrowth} are always interpolations
between all four models, but many models appear under the $H/D=2.5$
limit.

The region under each curve is the unstable region.  At every point of
the figure, the actual (in)stability of the disk is of course
determined by the highest limit. Not surprisingly, the disk is
effectively stabilized by increasing the velocity dispersion or by
adding mass to the inert halo, or a combination of both. This
behaviour has already been reported for various other cases, such as
Kuzmin potentials (\cite{Sellwood:Ath}, \cite{Ath:Sellwood}) and
quadratic potentials (\cite{Vau_Dej}). Another feature, which was also
shown to be present in quadratic potentials, is that the slope of the
stability limits increases for increasing $m$ (at least for $m \ge
2$). For hot disks, the $m=2$ mode is the only unstable one, while at
the cool end, $m=4$ turned out to be the most unstable perturbation.
Note that the model I was found to be stable for all calculated
harmonics.

The growth rate for the most unstable $m=2$ mode is shown in fig.
\ref{figgrowth}, again as a function of $\sigma_{\rm cent}$ and $H/D$.
This picture clearly shows that the growth rate tends to increase in a
more than linear way as the halo mass decreases. There is also a
tendency that the contour lines become less dependent on the velocity
dispersion for large values of the growth rate. This can be understood
from the fact that varying the velocity dispersion re-distributes the
weights $W(p)$ of the pole positions, and does not change the overall
intensity (which is influenced by the factor $H/D$). When
$\Im(\omega)$ becomes large, the pole distribution is ``seen'' from a
big distance, and the internal positions become less important than
the overall intensity.

\begin{figure}
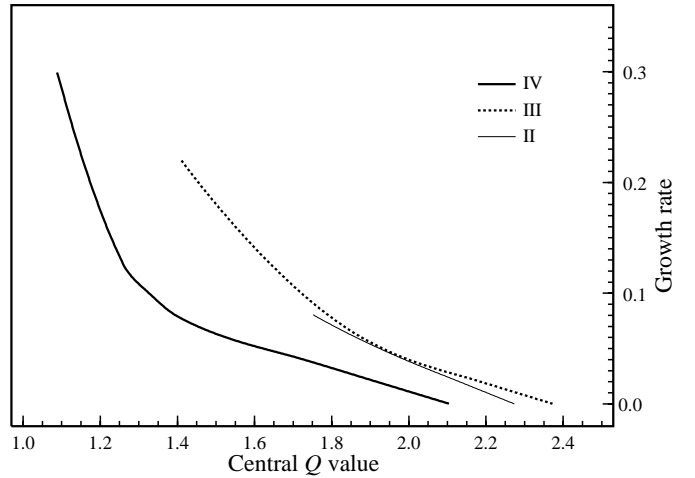

\vspace{7.0cm}
\special{ hscale=100 vscale=100 hoffset=-28 voffset= -30
hsize=480 vsize=500 angle=0 psfile=4583F13.ps } 
\caption{Growth rate of the $m=2$ mode as a \label{figToomre} 
function of Toomre's $Q$ factor in the centre of the galaxy.}
\end{figure}

Athanassoula and Sellwood (1986) suggest that a value of $2.0$-$2.5$
for Toomre's local stability parameter $Q$ might by a useful
criterion for stability of the disk against global $m=2$
perturbations. For the present models, fig. \ref{figToomre} plots the
growth rate of the dominant $m=2$ mode against the corresponding
$Q_{\rm cent}$, the value in the centre of the galaxy.  As $Q$ is an
increasing function of the radius, $Q_{\rm cent}$ is a lower limit of any
averaging of $Q$ over the disk. For each model, the variation in $Q$
is obtained by changing $H/D$. Equation (\ref{defQ}) shows that $Q$
simply scales with $\rho_0$, which is influenced by $H/D$.  The figure
shows that a value between $2.0$ and $2.4$ is indeed again a good
estimation of the stability limit. Our values are slightly larger than
those of Sellwood \& Athanassoula, but this can be explained by the
fact that they used a softened gravity.

\begin{figure}
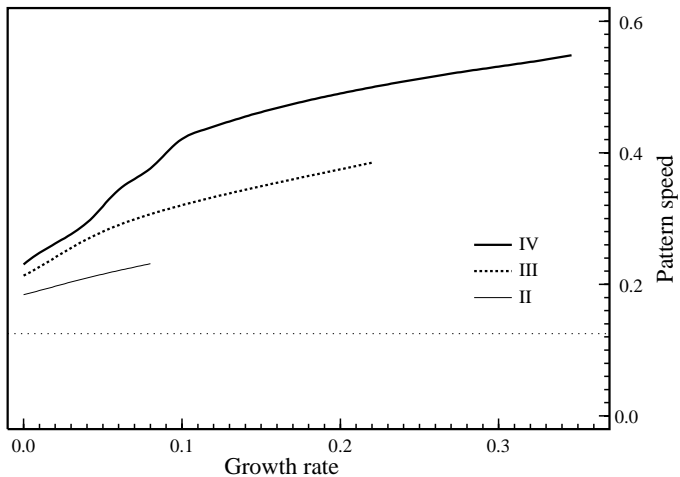

\vspace{7.0cm}
\special{ hscale=100 vscale=100 hoffset=-28 voffset=-30
hsize=480 vsize=500 angle=0 psfile=4583F14.ps } 
\caption{Pattern speed of the perturbation as a \label{figpattern}
function of the growth rate.  The dashed horizontal line corresponds
to the limit of the presence of an ILR.}
\end{figure}

The pattern speed of the perturbation, given by $\Re(\omega)/m$, is
plotted against the growth rate in fig. \ref{figpattern} (note that
$\Im(\omega)=0$ is, again, an extrapolation). This curve is shown for
the three unstable models II, III and IV, and the variation in the
growth rate is obtained by varying $H/D$. The endpoints of these
curves are a consequence of the self-consistency requirement. The
pattern speed has a clear increasing correlation with the growth rate,
which is again in agreement with N-body simulations
(\cite{Ath:Sellwood}). In addition, this figure shows that, for the
same growth rate, the pattern speed decreases with the velocity
dispersion. On the same picture, the upper limit on the pattern speed
for the presence of an ILR is shown as a horizontal line. From this,
it is clear that none of our models have an ILR, again in agreement
with the results decribed by Athanassoula \& Sellwood (1986).
In fact, orbits which are in ILR cause such a violent response (see
e.g. fig.
\ref{figpoles} and fig. \ref{figeigval}) that it would be very hard
for a galaxy to maintain much of them.

\begin{figure}
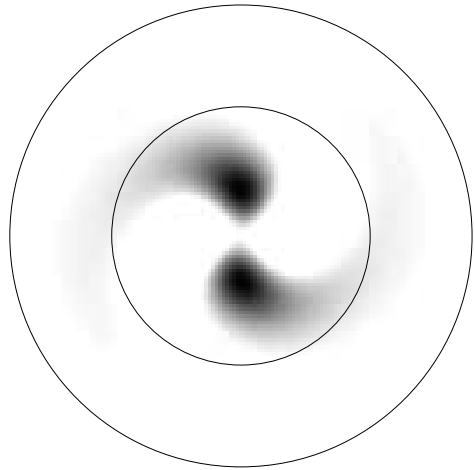

\vspace{6.3cm}
\special{ hscale=100 vscale=100 hoffset=20 voffset=-10
hsize=480 vsize=500 angle=0 psfile=4583F15.ps } 
\caption{Dominating $m=2$ mode for the model III at $H/D=2.5$. 
\label{spiral} The two circles correspond to the CR and OLR.
Both the streaming velocity and the pattern speed are clockwise.}
\end{figure}

As a matter of illustration, fig. \ref{spiral} shows the density
profile of the dominating $m=2$ mode in model III. This perturbation
develops its usual behaviour, with a bar-shaped central part and
losely wound spiral arms in the outer regions. The spiral structure is
always found to be trailing.


\section{Conclusions.}
A big part of this article was devoted to the description of a method
for finding linear modes in stellar disks. The proposed strategy
heavily relies on existing techniques, such as the matrix method and
Fourier expansion along the unperturbed orbit, but differs from
previous approaches by the fact that everything is calculated in
ordinary coordinate and velocity space and that a numerical set of
potential-density pairs is used. With the proposed scheme, the full
perturbed distribution function is obtained with no extra
calculation costs. In this way we have shown that calculations in
coordinate space, although much less suited for theoretical
considerations, can offer a fast and flexible alternative to
action-angle variables when it comes to numerical computation of
normal modes.

This method was applied to a set of unperturbed disk models,
having a more or less realistic potential and an exponential mass
density. This disks are embedded in a spherical, inert halo in order
to obtain a self-consistent model. In agreement with various other
studies (e.g.
\cite{Ath:Sellwood}; \cite{Vau_Dej}), the calculations showed that
these disks can be stabilised by increasing the velocity dispersion
and/or the halo mass.  For almost isotropic velocity dispersions, a
$Q_{cent}$ value of $2.0-2.5$ turned out to be a reasonable stability
limit. 

Comparison of the stability of the present models with the behaviour
of disks embedded in quadratic potentials (\cite{Vau_Dej}) shows a
striking resemblance. Qualitatively, the stability behaviour of those
simple uniformly rotating systems shares all the features that we have
found for the more sophisticated models discussed in this paper. And,
to a certain degree, there is even a quantitative agreement. It seems
that, for some important aspects of the stability behaviour, the
structure of the unperturbed distribution might be more inportant than
the nature of the unperturbed potential.


\acknowledgements{
The authors wish to thank the referee, C. Hunter, for a thorough
reading of the manuscript and the many valuable comments. P.
Vauterin acknowledges support of the Nationaal Fonds voor
Wetenschappelijk Onderzoek (Belgium)}


{}


\appendix

\section{The unperturbed distribution functions.}

\begin{figure}
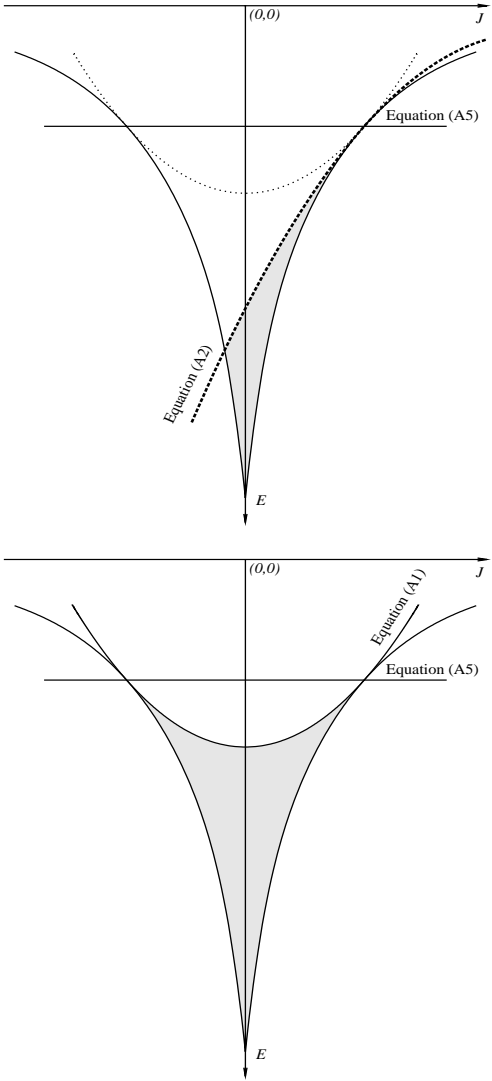

\vspace{14.8cm}
\special{ hscale=85 vscale=78 hoffset=-25 voffset=-150
hsize=480 vsize=500 angle=0 psfile=4583F16.ps } 
\caption{Representation of the orbits in $E$--$J$ space.
The shaded region contains the allowed orbits. The functional
dependence of the two limiting curves is given in the
text.\label{lind1} Top: limits used for the present models.
Bottom: simple $r_+ \le r_{\rm max}$ limits.}
\end{figure}

The simple $r_+=r_{max}$ limit of the unperturbed
distribution in phase space, defined in $E$-$J$ space by
\begin{equation}
E_L(E,J) = E+{J^2 \over 2 r_{max}^2} -V_0(r_{max}) \ge 0.
\end{equation}
is not so well suited for rotating models, since it is completely
symmetric with respect to $J$. Therefore we have chosen for an
alternative limit with the same zeroth and first derivative as $E_L$
at the point of the circular orbits with $r=r_{max}$ and with positive
$J$, but with a second derivative which can be chosen freely. It is
defined by
\begin{eqnarray}
E_{L,\beta} (E,J) = E-E_{c,max}
  &+& {J_{c,max} \over r_{max}^2} (J-J_{c,max}) \nonumber \\
  &-& \beta (J-J_{c,max})^2 \ge 0,\label{limi1}
\end{eqnarray}
with the binding energy for circular orbits with positive $J$ at
$r_{max}$
\begin{equation}
E_{c,max}=V_0(r_{max}) + {1 \over 2} r_{max} {d V_0 \over d r} (r_{max}),
\label{limi2}
\end{equation}
and the angular momentum at the same point
\begin{equation}
J_{c,max}= r_{max} \sqrt{
- r_{max} {d V_0 \over d r} (r_{max}) }.
\end{equation}
The parameter $\beta$ is adjustable.  For large $r$, this new limit
lies closely to circular orbits with positive $r$, and is much better
suited for rotating models. In order to avoid unwanted orbits with
$r_+ > r_{max}$, both limits (\ref{limi1}) and (\ref{limi2}) should be
combined with the extra condition (see also figure \ref{lind1})
\begin{equation}
E \le E_{c,max}.
\end{equation}

The unperturbed models are defined as
\begin{eqnarray}
\df_{\rm I}= E_{L,0.0} \times (\qquad
    3.84 \times 10^{-3} &.& E^4 \nonumber \\ 
  + 2.22 \times 10^{-1} &.& 1
\qquad),
\end{eqnarray}
\begin{eqnarray}
\df_{\rm II}= E_{L,0.007} \times (\qquad
    3.68 \times 10^{-2} &.& E^4 \nonumber \\
  + 3.15 \times 10^{-1} &.& E^4 \pw(J,2)
\qquad ),
\end{eqnarray}
\begin{eqnarray}
\df_{\rm III}= E_{L,0.01} \times (\qquad
    5.69 \times 10^{-4} &.& E^{11} e^{4 J} \nonumber \\
  + 7.69 \times 10^{-2} &.& E^8 \pw(J,2) \nonumber \\
  + 1.40 \times 10^{-1} &.& E^6 \pw(J,4)
\qquad),
\end{eqnarray}
\begin{eqnarray}
\df_{\rm IV}= E_{L,0.01} \times (\qquad
    6.69 \times 10^{-8} &.&  E^{25} e^{9 J} \nonumber \\
  + 7.90 \times 10^{-6}  &.& E^{25} \pw(J,2) \nonumber \\
  + 7.54 \times 10^{-3}  &.& E^{20} \pw(J,5) \nonumber \\
  + 4.29 \times 10^{-2}  &.& E^{14} \pw(J,7)
\qquad).
\end{eqnarray}
The function $\pw(x,n)$ which we introduced equals $x^n$ for $x>0$ and
zero for $x \le 0$. If $n$ is integer and larger than one, this
function has a continuous first derivative.

\end{document}